\documentstyle[epsfig]{mn}

\begin{document}
\def\ltsima{$\; \buildrel < \over \sim \;$}
\def\simlt{\lower.5ex\hbox{\ltsima}}
\def\gtsima{$\; \buildrel > \over \sim \;$}
\def\simgt{\lower.5ex\hbox{\gtsima}}
\def\approxgt{\mathrel{\hbox{\rlap{\lower.55ex \hbox {$\sim$}}
        \kern-.3em \raise.4ex \hbox{$>$}}}}
\def\approxlt{\mathrel{\hbox{\rlap{\lower.55ex \hbox {$\sim$}}
        \kern-.3em \raise.4ex \hbox{$<$}}}}

\title[Compton-thick Seyfert 2 galaxies with XMM-Newton]
{On the transmission- to reprocessing-dominated spectral state transitions
in Seyfert~2 galaxies}

\author[M.Guainazzi et al.]
{M.Guainazzi$^1$, A.C.Fabian$^2$, K.Iwasawa$^2$, G.Matt$^3$, F.Fiore$^4$\\ ~ \\
$^1$XMM-Newton Science Operations Center, European Space Astronomy Center, ESA, Apartado 50727, E-28080 Madrid, Spain\\
$^2$Institute of Astronomy, Madingley Road, Cambridge, CB3 0HA
\\
$^3$Dipartimento di Fisica ``E.Amaldi", Universit\`a ``Roma Tre", Via della Vasca Navale 84, I-00146 Roma, Italy 
\\
$^4$INAF-Osservatorio Astronomico di Roma, Via di Frascati 33, I-00040, Monteporzio, Italy
}

\maketitle
\begin{abstract}

We present {\it Chandra} and XMM-Newton observations
of a small sample (11 objects) of optically-selected Seyfert~2 galaxies,
for which ASCA and BeppoSAX had suggested Compton-thick
obscuration of the Active Nucleus (AGN). The main goal of this study
is to estimate the rate of transitions between ``transmission-''
and ``reprocessing-dominated'' states. We discover
one new transition in NGC~4939, with a possible
additional candidate in NGC~5643. This indicates a typical occurrence
rate of at least
$\sim$0.02~years$^{-1}$. These transitions could be
due to large changes of the obscuring gas column density,
or to a transient dimming of the AGN activity, the latter scenario
being supported by detailed analysis of the best
studied events.
Independently of the ultimate mechanism,
comparison of the observed spectral dynamics with Monte-Carlo simulations
demonstrates that the obscuring gas is largely inhomogeneous,
with multiple absorbing components possibly spread through the
whole range of
distances from the nucleus
between a fraction of parsecs up to several hundreds parsecs.
As a by-product of this study, we report the first measurement
ever of the column density covering the AGN in NGC~3393
($N_H \simeq 4.4 \times 10^{24}$~cm$^{-2}$), and
the discovery of soft X-ray extended emission, apparently
aligned along the host galaxy main axis in NGC~5005. The
latter object hosts most likely an historically misclassified
low-luminosity Compton-thin AGN.
 
\end{abstract}
\vspace{1.0cm}

\section{Introduction}

In X-rays,
obscured AGN may be classified into {\it Compton-thin} and {\it
Compton-thick}, according to the column of absorbing matter covering
the active nucleus. The threshold corresponds to a column density
$N_H \simeq \sigma_t^{-1} \simeq 1.5 \times 10^{24}$~cm$^{-2}$.
The fact that Compton-thick Seyfert~2s
are a substantial fraction of the whole population of Seyfert~2 galaxies,
maybe as high as 50\% (Risaliti et al. 1999), suggests that the
covering fraction of the absorbing matter is large.
If
a single absorber covers a steady-state active nucleus, 
the classification
of individual objects
is not expected to be time-dependent. A review on
the observational properties of Compton-thick Seyfert~2
galaxies has been recently published by Comastri (2004).

{\it Bona fide} Compton-thick Seyfert~2 galaxies are
observed in X-rays
also at energies lower than the photoelectric
cut-off. This X-ray emission  is probably due to
reprocessing of the nuclear emission by Compton-thick
matter surrounding the nucleus \cite{matt00b}, and/or by hot
plasma in the nuclear environment \cite{kinkhabwala02}.
We define hereafter {\it reprocessing-dominated}
Seyfert~2 galaxies those, whose X-ray emission in the
XMM-Newton energy band ($E \le 15$~keV) is dominated
by reprocessing\footnote{This definition is
therefore conceptually different from
{\it (Compton) reflection-dominated}
Seyfert~2s, where the emission in the XMM-Newton
energy band is dominated by Compton-reflection off
the far inner side of the absorber. Nevertheless,
almost all known ``reprocessing-dominated''
AGN are ``Compton reflection-dominated''.}.
The common wisdom so far has been to
{\bf identify reprocessing-dominated
Seyferts with Compton-thick AGN}.
However, very recently
transitions between ``Compton-thin''
and ``Compton-thick'' spectral states have
been serendipitously discovered in a few
X-ray bright Seyfert~2 galaxies
(Matt et al. 2003b, and references therein).
In UGC~4203, for instance (Guainazzi et al. 2001; Ohno et
al. 2004),
an XMM-Newton observation
detected a bright
(2--10~keV flux $\simeq 9 \times 10^{-12}$~erg~cm$^{-2}$~s$^{-1}$)
AGN, with a low-energy photoelectric cutoff
(corresponding to $N_H \simeq 2 \times 10^{23}$~cm$^{-2}$).
In ASCA observations, performed about six years earlier,
the weaker continuum and the
huge K$_{\alpha}$ fluorescent
iron line (Equivalent Width, $EW \simeq 1$~keV) can be
instead best explained
if the spectrum is dominated by the Compton echo
of an otherwise invisible nuclear emission.
Such transitions have been observed in both
directions, and are normally
accompanied by substantial changes
in the observed 2--10~keV flux.

This discovery stimulates some fundamental questions on
the nature of reprocessing-dominated Seyfert~2
galaxies.
These transitions could be due
in principle to a change of
the intervening absorption. 
Alternatively, Seyfert~2
X-ray spectral states dominated by reprocessing
may represent phases of low- or totally absent
activity in the life of an active nucleus, as observed,
for instance, in NGC~4051 (Guainazzi et al. 1998),
NGC~2992 (Gilli et al. 2000), and NGC~6300 (Guainazzi 2002).
In these cases, the observed transitions require
a change by at least one order of magnitude
of the nuclear activity level.

Transitions between ``Compton-thin'' and ``Compton-thick''
spectral states have been observed in 4 Seyfert 2 galaxies
so far (see Matt et al. 2003b, and references therein),
out of about 40 objects for which multiple X-ray
spectroscopic measurements are available (Bassani et al.
1999; Risaliti et al. 2001).
However, the ``parent sample''
is neither homogeneous, nor complete, being
substantially biased toward brighter (and therefore
less absorbed) objects (see the discussion in
Risaliti et al. 1999).

We are carrying on a XMM-Newton survey of
an optically defined and
complete - albeit small - sample of 
Seyfert galaxies,
classified as Compton-thick according to
observations prior to the launch of
{\it Chandra} and XMM-Newton.
The primary goal of this study is to determine the rate of
``transmission-" ({\it i.e.} Compton-thin)
to ``reprocessing-dominated" transitions\footnote{Although
in this paper we will refer to ``transmission-" {\it to}
``reprocessing-dominated" transitions, we
search for transitions in both directions},
and their typical timescale
on the soundest possible statistical basis.
This rate might be related to
the duty-cycle of the Active Galactic Nuclei
(AGN) phenomenon, at least in the
local universe, if these transitions
are due to large changes of
the overall X-ray AGN energy output \cite{matt03b}.
The results of this survey are the main subject of this
paper.

\section{The sample}

Our objects
are extracted from the sample of Risaliti et al. (1999), which
includes nearby Seyfert~2 galaxies with ASCA/BeppoSAX
measurements of the X-ray column density. We have restricted
our analysis to those objects, whose [O{\sc iii}] luminosity
is $> 10^{-13}$~erg~s$^{-1}$. As Risaliti et al.
(1999) discuss, this choice minimizes any bias due to incompleteness.
Out of the potential fourteen members of our sample,
priority ``A'' or ``B'' XMM-Newton observing time has
not been allocated to five of them (IC~2560; IRAS~07145; NGC~5135;
IC~3639; UGC~2456). We complement the XMM-Newton
observations with two objects observed by {\it Chandra}
(IC~2560; NGC~5135), whose data are available in the public
archive.
The galaxies discussed
in this paper are listed in Table~\ref{tab1}.
%-----------------Table 1
\begin{table*}
\caption{Sample discussed in this paper. $N_H$ measurements in this Table
were
performed prior to {\it Chandra} and XMM-Newton, and are listed in
Risaliti et al. (1999), except for the measurement in NGC~4945, which
is taken from Guainazzi et al. (2000a)}
\begin{tabular}{lccccccc} \hline \hline
Object & z & $N_{H,Gal}$ & $F_{[OIII]}$$^b$ & $N_H$$^a$ & XMM-Newton & Exposure time & Time span$^c$ \\
& & ($10^{20}$~cm$^{-2}$) & & (cm$^{-2}$) & Obs. date & pn/MOS or ACIS (ks) & (years) \\ \hline
NGC~1068 & 0.004 & 3.5 & 1580 & $>10^{25}$ & 29/30-Jul-2000 & 61.6/66.8 & 2.5  \\
Circinus & 0.0015 & 56 & 697 & $(4.3 \pm^{1.9}_{1.1}) \times 10^{24}$ & 6/7-Aug-2001 & 70.0/76.0 & 3.5 \\
NGC~5643 & 0.004 & 8.3 & 69 & $>10^{25}$ & 8-Feb-2002 & 7.1/9.4 & 4.9\\
NGC~1386 & 0.003 & 1.4 & 66 & $>10^{24}$ & 29-Dec-2002 & 13.6/17.0 & 6.0 \\
NGC~5135 & 0.014 & 4.6 & 61 & $>10^{24}$ & 4-Sep-2001$^d$ & 29.3 & 6.6 \\
NGC~3393 & 0.013 & 6.0 & 32 & $>10^{25}$ & 5-Jul-2003 & 10.9/14.2 & 6.5 \\
NGC~2273 & 0.006 & 7.0 & 28 & $>10^{25}$ & 5-Sep-2003 & 10.0/12.6 & 6.5 \\
NGC~5005 & 0.003 & 1.1 & 20 & $>10^{24}$ & 13-Dec-2002 & 13.1/8.8 & 7.0 \\
NGC~4939 & 0.010 & 3.4 & 11 & $>10^{25}$ & 03-Jan-2002 & 11.5/- & 5.0 \\
IC~2560 & 0.010 & 6.5 & $>4$ & $>10^{24}$ & 29/30-Oct-2000$^d$& 9.8 & 3.9 \\
NGC~4945 & 0.002 & 15.7 & $>4$ & $(4.4 \pm^{0.8}_{0.6}) \times 10^{24}$ & 21-Jan-2001 & 19.2/22.2 & 1.5 \\
\hline \hline
\end{tabular}

\noindent
$^a$after Risaliti et al. (1999); derived from ASCA or BeppoSAX observation

\noindent
$^b$in units of $10^{-13}$~erg~cm$^{-2}$~s$^{-1}$

\noindent
$^c$minimum distance between the ASCA/BeppoSAX and the {\it Chandra}/XMM-Newton
observation

\noindent
$^d${\it Chandra} observation

\label{tab1}
\end{table*}
%-----------------Table 1
EPIC spectra of four of the objects listed in Table~\ref{tab1}
have already been
individually published: NGC~1068 \cite{matt04},
the Circinus Galaxy \cite{molendi03}, NGC~5643
\cite{guainazzi04a}, NGC~4945 \cite{schurch02}.
The {\it Chandra} observation of IC~2560 is presented
by Iwasawa et al. (2002); the {\it Chandra}
observation
of NGC~5135 is discussed by Levenson et al. (2002, 2004).
For the
remaining five sources (NGC~1386, NGC~3393, NGC~2273,
NGC~5005, NGC~4939) we present here for the
first time the results of their XMM-Newton observations.

The average distance between the {\it Chandra}/XMM-Newton observation
and the latest ASCA/BeppoSAX one of the same object
is $\simeq$4.9~years.

In this paper: energies are quoted in the source
reference frame; errors on the count rate are
at the 1$\sigma$ level; uncertainties on the spectral
parameters are at the 90\% confidence level
for one interesting parameter; upper limits are as
well at the 90\% confidence level; in the
calculation of the luminosities, we adopted
a Hubble constant of 70~km~s$^{-1}$~Mpc$^{-1}$
\cite{bennett03}. Preliminary
results of this study are discussed by
Guainazzi et al. (2004b).

\section{Data reduction and analysis}

XMM-Newton data described in this paper
were reduced with SAS v5.4.1
\cite{jansen01}, using the most updated calibration
files available at the moment the data reduction
was performed. In this paper, only data from the
EPIC cameras (MOS; Turner et al. 2001; pn,
Str\"uder et al. 2001) will be discussed.
X-ray images are generally point-like.
Deviations from point-like
shapes are apparent in
NGC~1068 \cite{matt04}, the
Circinus Galaxy \cite{molendi03},
NGC~4945 \cite{schurch02}, NGC~5005
(Sect.~5).
Event lists from the two MOS cameras were
merged before accumulation of any scientific
products. Single to double (quadruple)
events were used to accumulate pn (MOS) spectra.
High-background particle flares were removed, by
applying standard thresholds on the single-event,
$E>10$~keV, $\Delta t = 10$~s light curves:
1~counts~s$^{-1}$ and 0.35~counts~s$^{-1}$ for each pn
and MOS camera, respectively.
Source spectra were extracted from
40$\arcsec$ circular regions around the X-ray nuclear source
centroid, except for NGC~5643 \cite{guainazzi04a},
where a smaller region was chosen to avoid a serendipitous
nearby bright source. Background scientific
products were extracted from annuli around the source
for the MOS, and circular regions in the same or nearby chips
for the pn, at the same height in detector coordinate as the
source location.
No significant variations in any energy
bands has been observed during the
XMM-Newton observations presented
here for the first time. Spectra were binned
in order to oversample the intrinsic
instrumental energy resolution by a factor
$\ge$3, and to have at least 25 counts in each
background-subtracted spectral channel. This
ensures that the $\chi^2$ statistics can
be used to evaluate the quality of the
spectral fitting. pn (MOS) spectra were
fitted in the 0.35--15~keV (0.5--10~keV)
spectral range.

\begin{figure*}
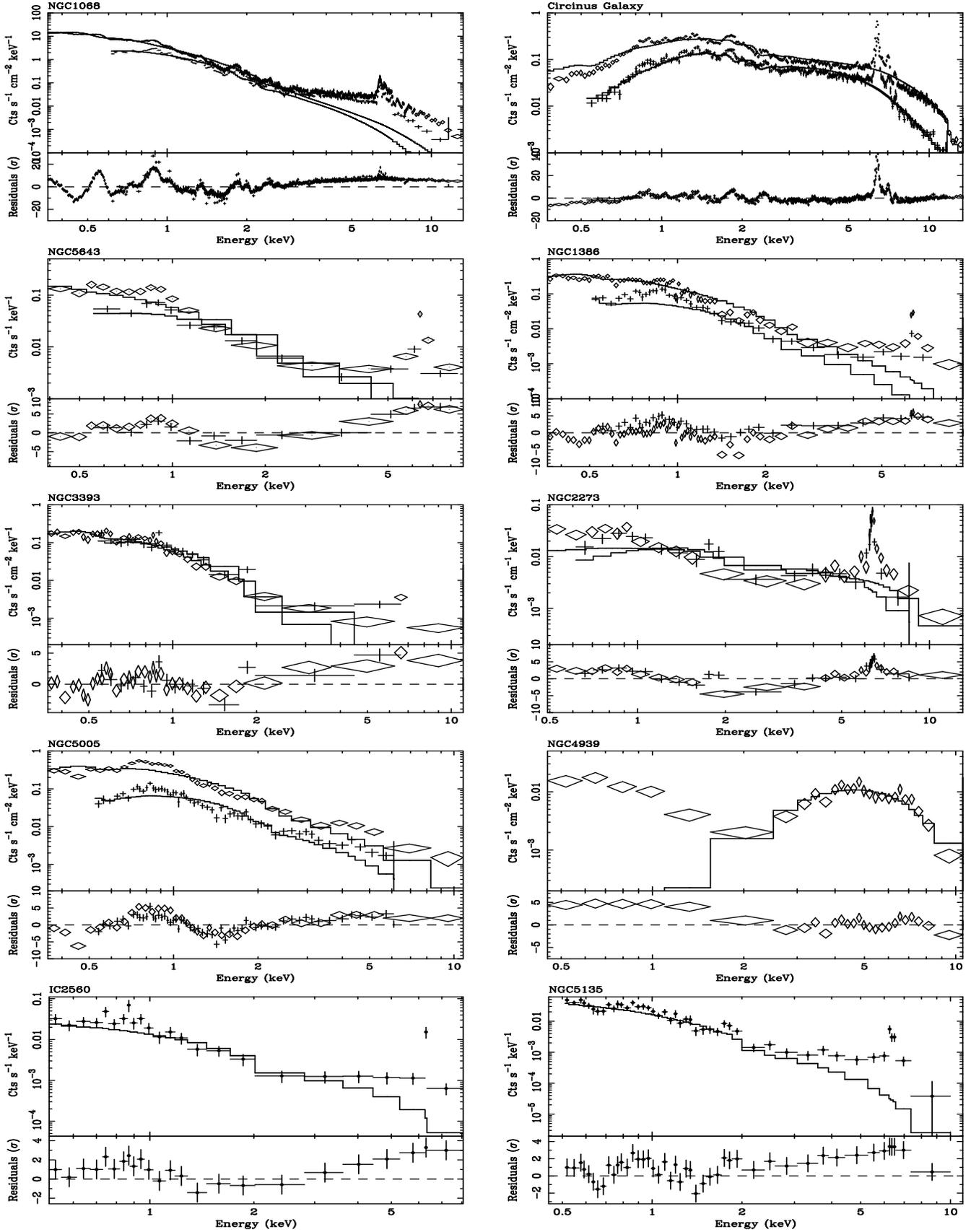

\hbox{
\hspace{-0.25cm}
\psfig{file=fig2a.ps,height=85mm,width=45mm,angle=-90}
\hspace{0.5cm}
\psfig{file=fig2b.ps,height=85mm,width=45mm,angle=-90}
}
\hbox{
\hspace{-0.25cm}
\psfig{file=fig2c.ps,height=85mm,width=45mm,angle=-90}
\hspace{0.5cm}
\psfig{file=fig2d.ps,height=85mm,width=45mm,angle=-90}
}
\hbox{
\hspace{-0.25cm}
\psfig{file=fig2e.ps,height=85mm,width=45mm,angle=-90}
\hspace{0.5cm}
\psfig{file=fig2f.ps,height=85mm,width=45mm,angle=-90}
}
\hbox{
\hspace{-0.25cm}
\psfig{file=fig2g.ps,height=85mm,width=45mm,angle=-90}
\hspace{0.5cm}
\psfig{file=fig2h.ps,height=85mm,width=45mm,angle=-90}
}
\hbox{
\hspace{-0.25cm}
\psfig{file=fig2i.ps,height=85mm,width=45mm,angle=-90}
\hspace{0.5cm}
\psfig{file=fig2j.ps,height=85mm,width=45mm,angle=-90}
}
\caption{Spectra ({\it upper panels}) and
residuals against a power-law continuum
modified by photoelectric absorption ({\it lower panels})
for the XMM-Newton and {\it Chandra}
observations of our sample. Readers are referred to
Schurch et al. (2002) for NGC~4945.
{\it Crosses}: MOS: {\it diamonds}: pn; {\it dots}: ACIS}
\label{fig2}
\end{figure*}

\begin{figure*}
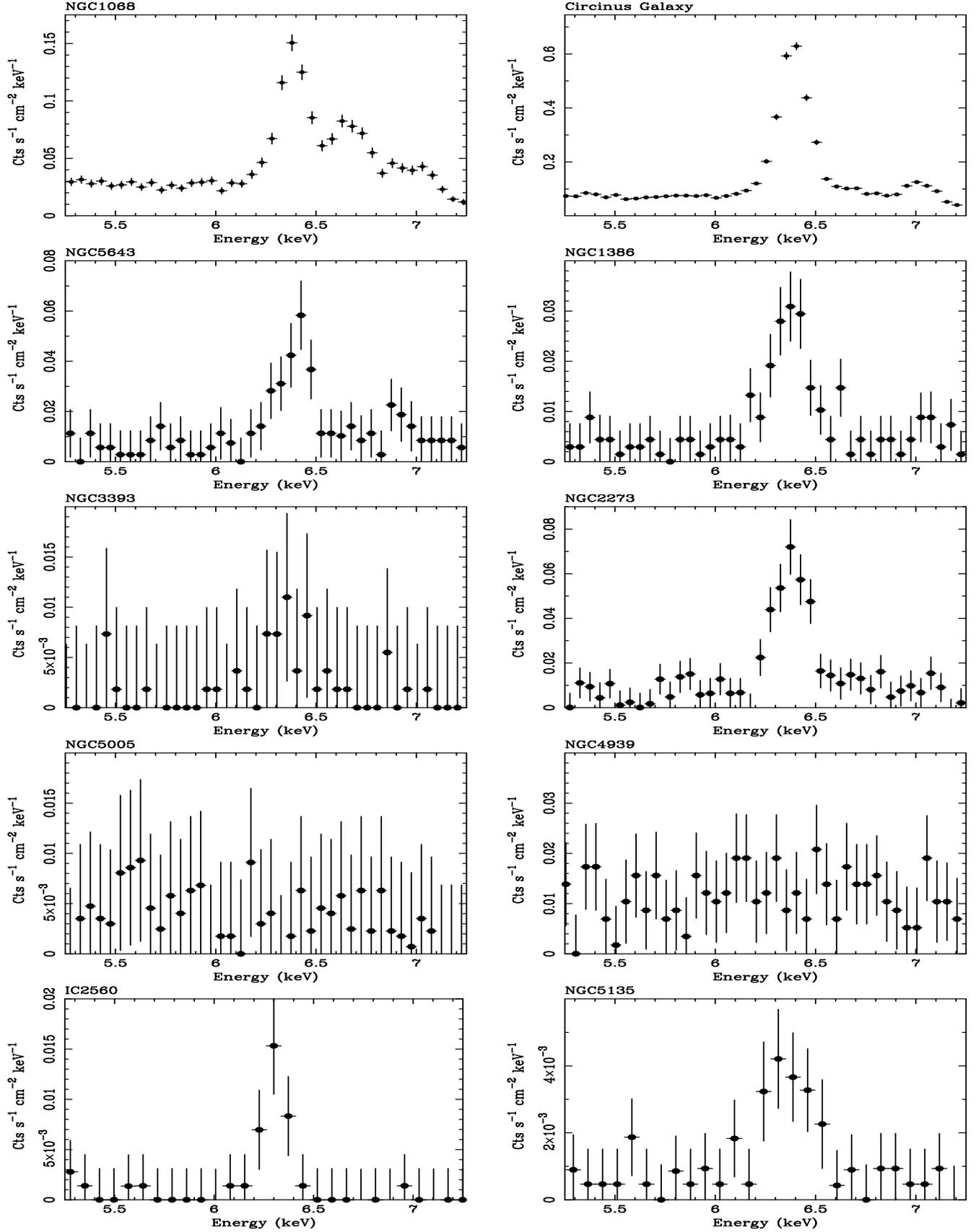

\hbox{
\hspace{-0.25cm}
\psfig{file=fig3a.ps,height=85mm,width=45mm,angle=-90}
\hspace{0.5cm}
\psfig{file=fig3b.ps,height=85mm,width=45mm,angle=-90}
}
\hbox{
\hspace{-0.25cm}
\psfig{file=fig3c.ps,height=85mm,width=45mm,angle=-90}
\hspace{0.5cm}
\psfig{file=fig3d.ps,height=85mm,width=45mm,angle=-90}
}
\hbox{
\hspace{-0.25cm}
\psfig{file=fig3e.ps,height=85mm,width=45mm,angle=-90}
\hspace{0.5cm}
\psfig{file=fig3f.ps,height=85mm,width=45mm,angle=-90}
}
\hbox{
\hspace{-0.25cm}
\psfig{file=fig3g.ps,height=85mm,width=45mm,angle=-90}
\hspace{0.5cm}
\psfig{file=fig3h.ps,height=85mm,width=45mm,angle=-90}
}
\hbox{
\hspace{-0.25cm}
\psfig{file=fig3i.ps,height=85mm,width=45mm,angle=-90}
\hspace{0.5cm}
\psfig{file=fig3j.ps,height=85mm,width=45mm,angle=-90}
}
\caption{Background-subtracted, linearly-rebinned
pn and ACIS
spectra in the 5.25-7.25~keV energy range
for the same objects as in Fig.~\ref{fig2}}
\label{fig3}
\end{figure*}

The residuals of fits
against a power-law continuum
modified by photoelectric absorption
are shown in Fig.~\ref{fig2}
for all the sources presented in this paper
except NGC~4945 \cite{schurch02}. 
%---------------------------------------
%---------------------------------------
Notwithstanding differences, and
despite the large
dynamical range in observed flux, the residuals
exhibit a remarkably similar pattern. Two continuum
components can be distinguished, joining
at $\simeq$2~keV (the only exception
being the Circinus Galaxy, whose soft
X-ray spectrum is heavily absorbed by
intervening matter in the plane of our
Galaxy). Spectra with the
best statistics show emission-like
features in the 0.5--1.5~keV energy range
(the exceptions being in this case
IC~2560, NGC~2273, and NGC~4939, which have the lowest
signal-to-noise soft X-ray spectra).
Above 2~keV spectra are flat, and exhibit almost ubiquitously
intense emission line features around
6~keV (observer's frame), the only
exceptions being NGC~5005, and NGC~4939.
The latter feature is most
straightforwardly explained as
iron K$_{\alpha}$ fluorescence.
These lines can be better appreciated
in Fig.~\ref{fig3}, where we show
%---------------------------------------
%---------------------------------------
background-subtracted spectra
in the energy
range around the K$_{\alpha}$ iron line with
a constant linear binning of about 50~eV.

In Sect.~4 observed spectra will be
compared against composite ``two-continuum''
scenarios. In these scenarios,
the soft X-ray spectrum can
be accounted for by one of the possible
model combinations:

\begin{itemize}

\item emission from an optically thin,
collisionally ionized plasma ({\tt mekal} in
{\sc Xspec}, Mewe et al. 1985)
with free
elemental abundances ({\it ``thermal scenario''} hereafter)

\item a power-law with free spectral index $\Gamma_{soft}$,
plus as many unresolved emission lines as 
required according to a 90\% confidence level
F-test criterion ({\it ``scattering scenario''})

\end{itemize}

The hard X-ray continuum will be instead accounted
for by
one of the following models:

\begin{itemize}

\item a power-law
with free spectral index $\Gamma_{hard}$,
covered by photoelectric absorption with
column density
$N_H$ ({\it ``transmission scenario''})

\item a ``bare'' ({\it i.e.} unabsorbed)
Compton-reflection spectrum
({\tt pexrav} in {\sc Xspec}; Magdziarz \& Zdziarski 1995)
with solar abundances ({\it ``(Compton-)reflection scenario''})

\end{itemize}

These simple parameterizations yield adequate
fits for all the spectra presented in this
paper. One should, however, be aware of
possible limitations inherent to this simple
approach. High-resolution
spectroscopy of nearby Seyfert~2
galaxies (among which NGC~1068; Kinkhabwala et al. 2002;
Brinkman et al. 2002) has convincingly
demonstrated that soft X-ray emission is
dominated - at least in some cases -
by emission lines, with negligible contribution by an
underlying continuum. Blending of these
emission lines in the EPIC spectra can
mimic a continuum emission. This point
is discussed in larger extent by Iwasawa et al
(2002). As our primary
concern in this paper is the characterization of
the nuclear absorber, the uncertainties
induced by a purely phenomenological
modeling of the soft X-ray spectrum will
not substantially affect the core results
of our paper (Guainazzi et al. 2004a).
In the above modeling, we exclude moreover
the possibility that the
reprocessed component dominating
the hard X-ray spectrum in the ``reflection
scenario'' is in turn absorbed - {\it e.g.}
by the near side outer rim or atmosphere of the
same matter, responsible for reprocessing. This
possibility is discussed by Guainazzi et al.
(2004a) with respect to the NGC~5643 case.
In none of the other sources discussed
in this paper we have found convincing evidence
for this possibility. However, statistics
is often not good enough to strictly rule it out.

\section{XMM-Newton/{\it Chandra} results}

In this Section we summarize the results of the
XMM-Newton and {\it Chandra} (IC~2560 and NGC~5135)
observations
of the targets listed in Table~\ref{tab1}.

\subsection{NGC~1068}

NGC~1068 is one of the X-ray brightest and best studied
Compton-thick Seyfert~2 galaxies. Its Compton-thick
nature had been suggested by the prominent and
multi-component K$_{\alpha}$ emission line complex
observed by ASCA (Ueno et al. 1994; Iwasawa et al. 1997),
and finally confirmed by BeppoSAX \cite{matt97a}.
The column density of
the absorber covering the active nucleus
probably exceeds $10^{25}$~cm$^{-2}$ \cite{matt97a}.
The soft X-rays are dominated by
line emission following photoionization and
photoexcitation by the active nucleus emission
\cite{kinkhabwala02}, with little contribution from
the circumnuclear starburst \cite{wilson92}.

The EPIC spectrum of the XMM-Newton observation is
discussed by Matt et al. (2004). Several Fe and
Ni emission lines allowed them to study in  
details the nature of the reflecting matter.
Detection of iron K$_{\alpha}$ Compton-shoulder
confirms that the neutral reflector is Compton-thick.
It is
likely to be the far side inner wall of the
absorber. Iron (nickel) overabundance of a factor about 2 (4),
for lower Z elements when compared to
solar values
was measured as well.

\subsection{The Circinus Galaxy}

The Circinus Galaxy hosts the closest known active
nucleus. ASCA unveiled a reprocessing-dominated spectrum
\cite{matt96}. Detection of the nuclear emission in
the PDS instrument on-board BeppoSAX \cite{matt99}
allowed to precisely measure the column density of
the absorber covering the nucleus ($N_H \simeq 4 \times
10^{24}$~cm$^{-2}$). In hard X-rays the
nuclear emission is dominated by an unresolved bright
core on scales $<8$~pc \cite{sambruna01}. The EPIC
hard X-ray spectra are discussed by Molendi et al. (2003).
Again, the measurement of iron K$_{\alpha}$
Compton-shoulder -
previously discovered by {\it Chandra} \cite{bianchi02} -
allowed them to identify  matter responsible for the
Compton-reflection dominating
below 10~keV with the Compton-thick
absorber

\subsection{NGC~5643}

Maiolino et al. (1998) classified NGC~5643 as a
Compton-thick ($N_H > 10^{25}$~cm$^{-2}$) Seyfert~2
galaxy, whose 0.1--10~keV spectrum is dominated
by free electron scattering. However, in a later
XMM-Newton
pointing Guainazzi et al. (2004a) measured
a line-of-sight column density
in this object, comprised
between 0.6 and $1.0 \times 10^{24}$~cm$^{-2}$.
The absorber may be directly covering the nuclear
emission or its Compton-reflection. Comparison with
previous BeppoSAX and ASCA observations unveiled dramatic
changes in the 1--10~keV spectral shape, which can
be parameterized as an {\it observed} photon index
dynamical range $\Delta \Gamma \simeq 2.0$ accompanying
a variation of the 2--10~keV flux by a factor $>$10.
 The extreme variability
observed in the nuclear emission of this object
indicates the revival of an
AGN which was ``switched-off" during the BeppoSAX
observation.
The interpretation of this large variation is,
however, complicated by the fact that the large
ASCA and BeppoSAX apertures ($\simeq$3$^{\prime}$)
encompass a bright serendipitous source (christened
``NGC~5643 X-1" by Guainazzi et al. 2004a), apparently
located in one of the wide spiral arm of this
face-on galaxy. Understanding the
spectral dynamics associated with the
flux changes requires instruments
capable of distinguishing the contribution of the
two bright X-ray sources.

\subsection{NGC~5135}

We have reanalyzed the {\it Chandra} observation
already discussed by Levenson et al. (2004). Our results
are substantially coincident with theirs. The ACIS-S3
spectrum is best-fit in the ``thermal+reflection'' scenario.
The soft X-ray spectrum requires two thermal components
with $kT \sim 80$ and $\simeq 390$~eV, plus
an additional emission line with centroid energy
$E_c \simeq 1.78$~keV. Above 2~keV the spectrum is
Compton-reflection dominated, consistent with the
AGN being obscured
by a column density $N_H \approxgt 9 \times 10^{23}$~cm$^{-2}$
(for an intrinsic photon index of 1.5 and a reflection
fraction $\le$0.5). The intensity
of the K$_{\alpha}$ fluorescent emission line is
$(5.2 \pm^{1.9}_{2.6}) \times 10^{-6}$~photons~cm$^{-2}$~s$^{-1}$,
corresponding to an EW against the reflection continuum of
$1.7 \pm^{0.6}_{0.8}$~keV. The absorption-corrected fluxes
in the 0.5--2~ and 2--10~keV energy bands are $(1.9 \pm^{2.8}_{1.0})$
and $(1.6 \pm^{1.0}_{0.6}) \times 10^{-13}$~erg~cm$^{-2}$~s$^{-1}$,
respectively.

\subsection{NGC~1386}

\begin{figure*}
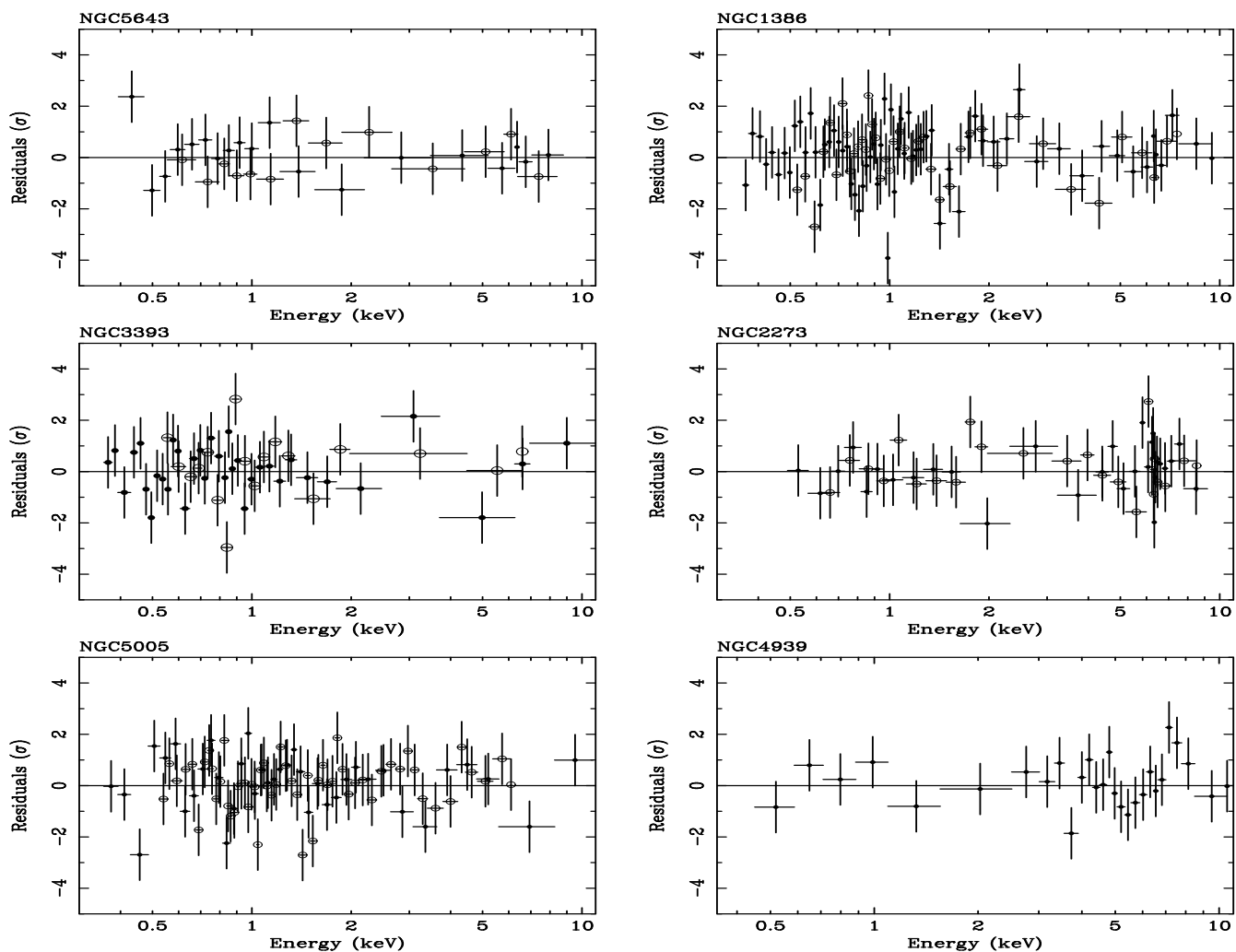

\hbox{
\hspace{-0.25cm}
\psfig{file=fig4c.ps,height=85mm,width=45mm,angle=-90}
\hspace{0.5cm}
\psfig{file=fig4d.ps,height=85mm,width=45mm,angle=-90}
}
\hbox{
\hspace{-0.25cm}
\psfig{file=fig4e.ps,height=85mm,width=45mm,angle=-90}
\hspace{0.5cm}
\psfig{file=fig4f.ps,height=85mm,width=45mm,angle=-90}
}
\hbox{
\hspace{-0.25cm}
\psfig{file=fig4g.ps,height=85mm,width=45mm,angle=-90}
\hspace{0.5cm}
\psfig{file=fig4h.ps,height=85mm,width=45mm,angle=-90}
}
\caption{Residuals in units of standard
deviations against the best-fit models
as in Table~\ref{tab2}, plus NGC~5643; pn: {\it filled dots};
MOS: {\it empty circles}.}
\label{fig4}
\end{figure*}

The results of the XMM-Newton observation of NGC~1386 are presented
for the first time in this paper. Two of the baseline scenarios
can be ruled out. The ``scattering+reflection" scenario can be
rejected, as it produces a rather bad $\chi^2/\nu = 172.7/83$.
The ``thermal+transmission" scenario yields a better fit
($\chi^2/\nu = 131.5/92$). However, it requires a rather flat
AGN spectral index ($\Gamma \simeq 0.5$). The two remaining
scenarios yield comparably good fits:
``scattering+transmission": $\chi^2/\nu = 135.4/84$;
``thermal+reflection": $\chi^2/\nu = 133.4/94$.
In the former, the EW of the K$_{\alpha}$ iron line
($EW \simeq 1.0$~keV) is too large with respect to
the expected values for transmission through a uniform
shell of material encompassing the
continuum source \cite{leahy93}, assuming the best-fit
$N_H \simeq 4 \times 10^{23}$~cm$^{-2}$.
In the latter, two thermal components are
required to account for
the bulk of the soft X-rays, alongside a Compton-reflection
component plus iron K$_{\alpha}$ iron line dominating
above about 2~keV. The best-fit parameters for the
fits discussed in this Section
are reported in Table~\ref{tab2}.
%------------------- Table 2
\begin{table*}
\caption{Best-fit parameters
and results for the sources in Table~\ref{tab1},
whose XMM-Newton EPIC spectral fitting results are presented
for the first time in this paper. The legenda for the
``Model'' column two-letters code
is as follows: the first letter indicates
the scenario, which best accounts for the soft X-ray
spectrum: scattering (``S''), or thermal emission
(``T''); the second letter indicates the scenario, which
best accounts for the hard X-ray spectrum: transmission
(``T''), or reflection (``R'')
}
\begin{tiny}
\begin{tabular}{lcccccccccc} \hline \hline
& &\multicolumn{2}{c}{Hard X-ray continuum} & \multicolumn{3}{c}{Emission lines} & \multicolumn{3}{c}{Soft X-rays} & \\
Source & Model & $\Gamma_{hard}$ & $N_H$$^a$ & $E_c$ & $I_c$$^b$ & $EW$ & $kT$ & $Z$ & $\Gamma_{soft}$ & $\chi^2/\nu$ \\
& & & ($10^{23}$~cm$^{-2}$) & (keV) & & (keV) & (keV) & ($Z_{\odot}$) & \\ \hline
NGC~1386 & TR &$2.5 \pm^{0.5}_{0.4}$ & $\ge 22$ & $6.41 \pm^{0.02}_{0.03}$ & $0.81 \pm^{0.16}_{0.14}$ & $1.8 \pm^{0.4}_{0.3}$ & $0.12 \pm^{0.05}_{0.02}$ & $0.07 \pm 0.02$ & ... &  133.4/94 \\ 
 & &  & &  & & & $0.66 \pm 0.03$ & $\equiv Z$(0.12~keV) &  &  \\ 
NGC~3393 & TR & $1.6 \pm 1.2$ & $\ge 9$ & $6.4^c$ & $0.25 \pm 0.14$ & $1.4 \pm 0.8$ & $0.14 \pm^{0.04}_{0.03}$ & $0.04 \pm ^{0.03}_{0.02}$ & ... &  55.8/43 \\
&  & & & $1.84 \pm^{0.12}_{0.04}$ & $0.15 \pm 0.09$ &  & $0.57 \pm^{0.06}_{0.08}$ & $\equiv Z$(0.14~keV)  &  &   \\
NGC~2273 & TR & $1.5 \pm 0.4$ & $\ge 18$ & $6.400 \pm 0.010$ & $2.3 \pm^{0.4}_{0.3}$ & $2.2 \pm^{0.4}_{0.3}$ & $0.8 \pm 0.2$ & $<0.06$ & ... &  55.2/51 \\
NGC~5005 & TT & $1.6 \pm^{0.7}_{0.6}$ & $0.3 \pm 0.2$ & 6.4$^c$ & $<0.14$ & $<0.24$ & $0.60 \pm^{0.03}_{0.02}$ & $0.30 \pm^{0.14}_{0.30} $ & ... & 126.4/122 \\
 & &  & &  & & & $2.3 \pm^{1.1}_{0.7}$ & $\equiv Z$(0.60~keV) &  &  \\ 
NGC~4939 & ST & $1.5 \pm 0.5$ & $1.5 \pm^{0.4}_{0.5}$ & 6.4$^c$ & $<0.4$ & $<0.07$ & ... & ...  & $2.7 \pm 0.4$ & 21.1/21 \\
 &  & & & $6.96$$^c$ & $<1.1$ & $<0.21$ & &  &  &   \\ 
\hline \hline
\end{tabular}
\end{tiny}

\noindent
$^a$calculated assuming $\Gamma$ frozen to its best-fit value for Compton-reflection dominated spectra, and a reflection fraction $\le 0.5$

\noindent
$^b$in units of $10^{-5}$~photons~cm$^{-2}$

\noindent
$^c$frozen

\label{tab2}
\end{table*}
%------------------- Table 2
Residuals against the best-fit models are shown in Fig.~\ref{fig4}.
%---------------------------------------
%---------------------------------------

\subsection{NGC~3393}

NGC~3393 is the object in our sample with the lowest signal-to-noise
in the hard X-ray band. The iron line is barely detectable
above a very weak continuum, with  $\Delta \chi^2/\Delta
\nu = 6.2/1$, corresponding the the 98.3\% confidence level,
if one assumes that the line is predominantly neutral..
The scattering scenario
yields $\chi^2_{\nu} \simeq 1.7$. Thermal
model for the soft X-ray spectra
produces a significantly better fit. In the hard
X-ray band, transmission- and reflection-dominated scenarios yield
statistically comparable fits.
In the former scenario the
EW of the K$_{\alpha}$ iron line ($EW = 440 \pm 180$) is about one
order-of-magnitude larger than expected from the
measured column density
[$N_H = (7 \pm^7_4) \times 10^{22}$~cm$^{-2}$, if $\Gamma \equiv 1.9$].
We conclude therefore that
Compton-reflection dominance is the most plausible explanation
for the hard X-ray spectrum in this object.
The lower limit on the column density covering the
active nucleus derived from the XMM-Newton observation
[$N_H > 7 (9) \times 10^{23}$~cm$^{-2}$ if
$\Gamma = 1.6$~(1.9)]
strictly speaking does
not rule out an - albeit extreme - Compton-thin absorber.
Nonetheless, its ultimate nature is confirmed by
a reanalysis of the BeppoSAX observation
(cf. Sect.~5). An emission
line with centroid energy
$E_c \simeq 1.8$~keV is required at the
95.1\% confidence level
($\Delta \chi^2/\Delta \nu = 8.4/2$). This line
may correspond to K$_{\alpha}$ fluorescence of
Si, which is expected to be produced by
Compton-reflected spectra. However, its EW against the
reflected continuum is $\sim$5~keV, too large to
be produced by the same Compton-reflection responsible
for the iron emission \cite{matt97b}.

\subsection{NGC~2273}

For NGC~2273 the family of
models where hard X-rays are accounted for
by an absorbed power-law yield an unacceptably flat intrinsic
spectral index ($\Gamma \simeq -0.2$--0.5), as well as
an unacceptably large EW of the iron K$_{\alpha}$ iron line
($EW \simeq 2.3$--3.6~keV) with respect to the measured
column density ($N_H \simeq 1.4$--$12 \times 10^{22}$~cm$^{-2}$).
Compton-reflection domination is a viable alternative.
Modeling the soft X-rays with the ``thermal" or the ``scattering"
scenario makes very little difference on the properties
of the hard X-ray continuum or of the
K$_{\alpha}$ iron line, although in the latter scenario
the photon index best-fit
value is closer to standard values for
AGN ($\Gamma \simeq 1.5$ versus 1.2, respectively).
In Table~\ref{tab2} we list the results obtained with the former.

\subsection{NGC~5005}

\begin{figure*}
\hbox{
\psfig{file=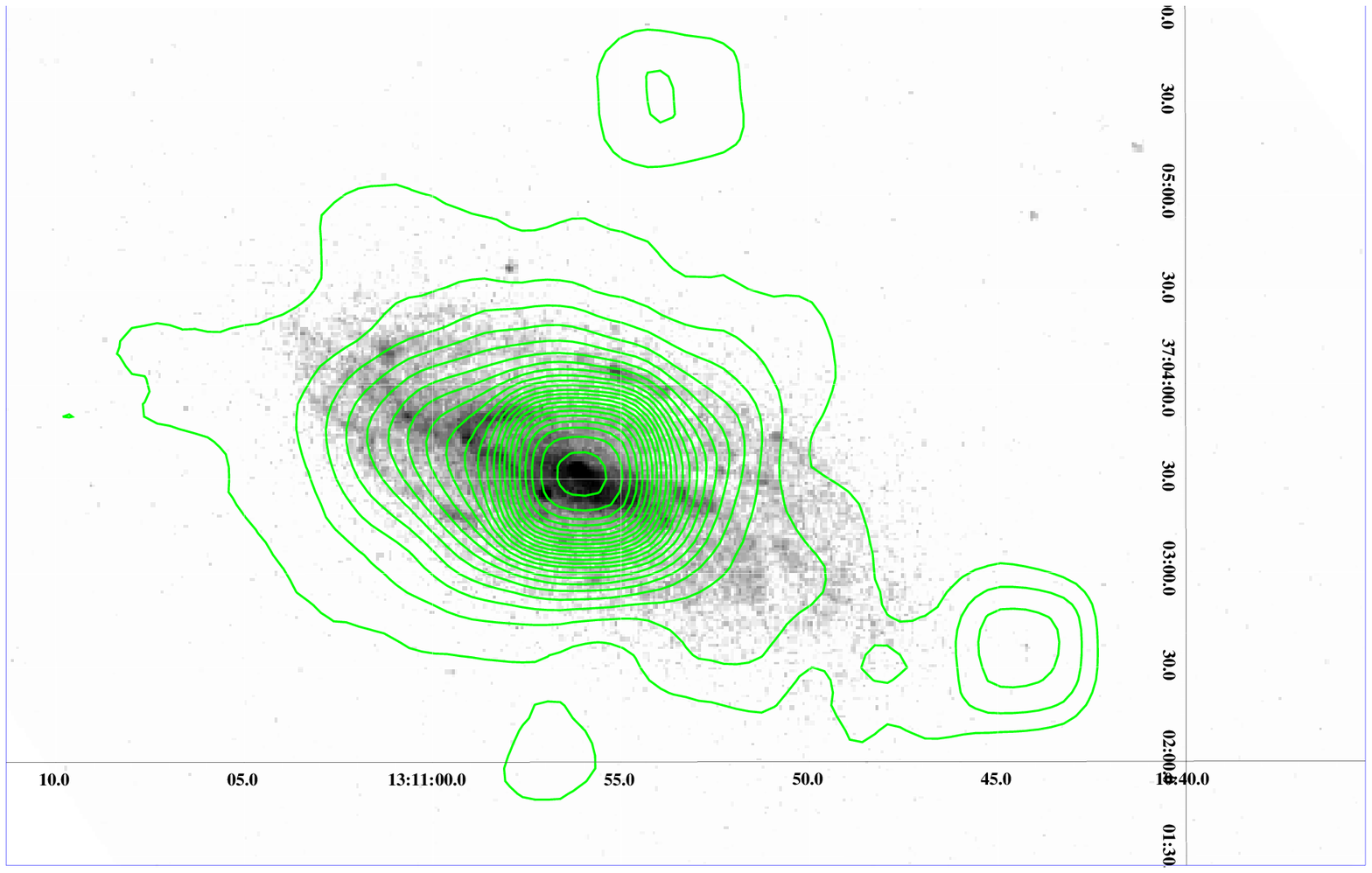,height=70mm,width=85mm}
\hspace{0.2cm}
\psfig{file=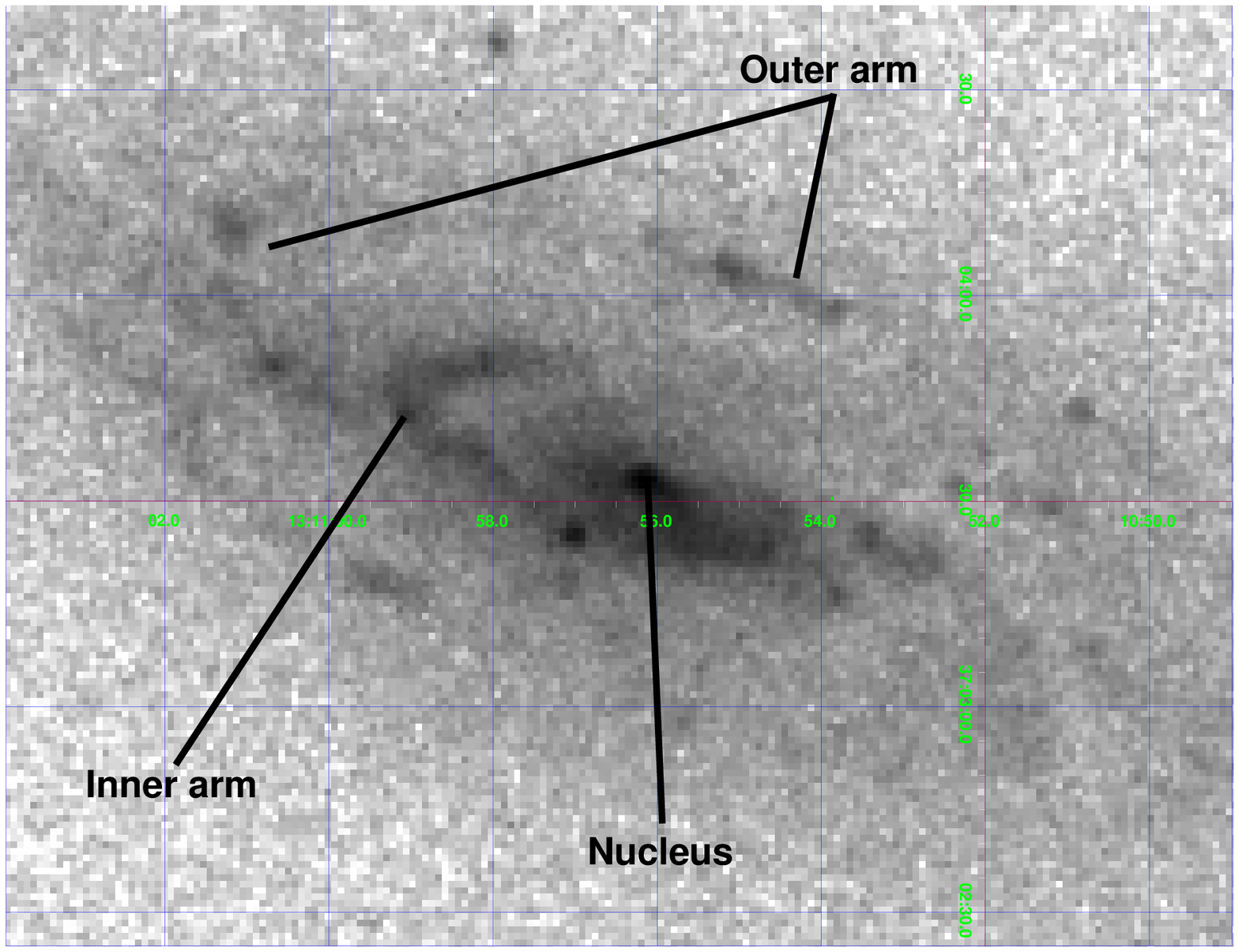,height=70mm,width=85mm}
}
\caption{{\it Left}: pn intensity contours in the 0.2--15~keV
energy band superposed to the OM exposure with the
UVW1 filter for NGC~5005. The pn image is smoothed
with a 6$^{\arcsec}$ kernel boxcar function.
Contours represent 15 logarithmically equispaced count
levels from 0.17 to 18.29 on the smoothed image.
{\it Right}: zoom of galaxy surface in the UVW1 filter.
The position of the nucleus, of the inner and the outer
arms are labeled}
\label{fig6}
\end{figure*}

The XMM-Newton observation shows that
the X-ray emission is extended,
and apparently elongated along a direction close to the main
axis of the host galaxy, or coincident
with an inner spiral arm,
visible in the simultaneous OM UVW1 filter
(2500-4000$\AA$) image
(Fig.~\ref{fig6}).
%---------------------------------------
%---------------------------------------
Although the diffuse emission is mostly
associated with soft X-rays, the
statistics is not good enough to estimate a
threshold energy, above which the X-ray emission is
no longer extended.
Assuming that the diffuse emission is
due to shocked gas in regions of intense star formation,
we have considered
only models
where at least part of the soft X-rays are due to
a thermal component. Hard X-ray Compton-dominance is
unlikely. A fit where the hard X-ray emission is due
to a ``bare" Compton-reflection yields a very steep
intrinsic spectral index ($\Gamma_{hard} \simeq 3.1$). Moreover,
no iron K$_{\alpha}$ fluorescent line is detected,
and the upper limit
on the EW of a 6.4~keV narrow Gaussian profile is
rather strict ($\le 240$~eV).
Transmission through a moderate absorber ($N_H \simeq
3 \times 10^{22}$~cm$^{-2}$) is a viable alternative.
The soft X-rays can be accounted for by the combination
of two thermal components ($\chi^2/\nu = 126.4/122$) or
of one thermal component and a scattered power-law 
($\chi^2/\nu = 138.9/124$). In Tab.~\ref{tab2} we show
the results obtained in the former scenario. In the
latter, $\Gamma_{hard} \simeq 1.8$, and
$N_H \simeq 5 \times 10^{22}$~cm$^{-2}$.

\subsection{NGC~4939}

NGC~4939 was serendipitously
located in the pn field of view of
an observation of SAX~J1305.2-1020. Its spectrum
is the only one
in our sample, which clearly exhibits a soft
photoelectric cut-off (cf. Fig.~\ref{fig2}).
Indeed, the ``transmission" scenario accounts
well for the hard X-rays, with
$N_H \simeq 1.5 \times 10^{23}$~cm$^{-2}$. Modeling
the soft X-rays with a single thermal component
($kT \simeq 0.7$~keV) or a steep power-law
($\Gamma \simeq 2.7$) yields fits of equivalent
statistical quality: $\chi^2/\nu = 18.9/20$,
and 21.1/21, respectively. The best-fit
parameters and results for the latter are shown
in Table~\ref{tab2}. An emission line is additionally
required at the 94.0\% confidence level only
($\Delta \chi^2/\Delta \nu = 5.4/2$). Its
centroid energy is inconsistent with emission
from neutral iron: $E_c = 6.71 \pm^{0.12}_{0.20}$~keV.
If, following Maiolino et al. (1998; cf. Sect.~5 as well),
we interpret this shift of the centroid energy as due to
a blend of neutral and a H-like transitions,
the 90\% upper limits on the EW of
either component are 70~eV and
210~eV, respectively.

\subsection{NGC~4945}

The XMM-Newton observation of NGC~4945 is discussed
by Schurch et al. (2002). The galaxy core has a
complex morphology. It is dominated by reprocessing, as
the nucleus is covered by a thick absorber
($N_H \simeq 4 \times 10^{24}$~cm$^{-2}$; Done et al. 1996).
Compton-reflection from the inner side of an edge-on
torus leaves its imprinting in the
hard X-ray spectrum through a 1.6~keV EW
K$_{\alpha}$ emission line, consistent with previous
findings (Guainazzi et al. 2000a; Madejski et al. 2000).
In soft X-rays,
multi-temperature emission from a nuclear starburst
dominates, a two temperature model yielding
$kT \simeq$0.9~keV and $kT \simeq$~6.9~keV.
The hard X-ray emission exhibits
a resolved morphology, suggesting that
part of the gas in the starburst region is
exposed to the AGN radiation as well. 

\subsection{IC~2560}

IC~2560 has not been observed by XMM-Newton. The
results of a {\it Chandra} observation of this
target are discussed by Iwasawa et al. (2002).
A model constituted by a two-component thermal emission
plus a Compton reflection dominated spectrum
(with $\Gamma \equiv 1.9$) is an adequate
description of the spectrum ($\chi^2/\nu = 61.2/74$).
The EW of the K$_{\alpha}$ iron line is $\simeq 3.5$~keV.
In principle, a statistically equivalent fit is obtained
if the ``bare" Compton-reflection component is substituted
by an absorbed power-law ($\chi^2/\nu = 59.9/73$). However,
in this scenario the K$_{\alpha}$ iron line EW ($\simeq
8.0$~keV) is almost two orders of magnitude too large than expected
in transmission from the measured column density
($N_H \simeq 5 \times 10^{22}$~cm$^{-2}$). The
{\it Chandra} observation therefore supports the
interpretation of the
IC~2560 ASCA spectrum as hard
X-ray reprocessing-dominated \cite{risaliti00},
against the interpretation
of the same data
in terms of a Compton-thin absorber covering the nuclear
emission
given by Ishihara et al. (2001)

\subsection{Fluxes and luminosities}

In Table~\ref{tab5} we present the
%------------- Table 5
\begin{table}
\caption{Observed fluxes for the sources listed in
Table~\ref{tab2}. Units are in $10^{-12}$~erg~cm$^{-2}$~s$^{-1}$}
\begin{tabular}{lcc} \hline \hline
Source & 0.5--2~keV & 2--10~keV \\ \hline
NGC~1386 & $1.8 \pm^{0.9}_{0.5}$ & $0.27 \pm 0.05$ \\
NGC~3393 & $0.43 \pm^{0.39}_{0.18}$ & $0.09 \pm^{0.06}_{0.04}$ \\
NGC~2273 & $0.12 \pm^{0.18}_{0.06}$ & $0.69 \pm^{0.16}_{0.12}$ \\
NGC~5005 & $0.47 \pm 0.03$ & $0.51 \pm 0.06$ \\
NGC~4939 & $0.12 \pm 0.04$ & $3.3 \pm^{0.10}_{0.30}$ \\ \hline \hline
\end{tabular}
\label{tab5}
\end{table}
%------------- Table 5
observed fluxes in the 0.5--2~keV and
2--10~keV energy ranges for all the sources
in Table~\ref{tab2}. The corresponding
luminosity corrected for Galactic absorption
in the soft
X-ray band ranges between $10^{41}$ to
$10^{43}$~erg~s$^{-1}$ (cf. Fig.~\ref{fig11}). In the hard X-ray
band, the determination of the intrinsic
AGN luminosity is impossible for all cases where
only lower limits on the nuclear absorbing column
density exist. For Compton-thick sources
they are anyhow plagued by large uncertainties.
For the two sources which are Compton-thin
in XMM-Newton observations, the
2--10~keV luminosity is
$1.8 \times 10^{42}$~erg~s$^{-1}$ (NGC~4939),
and $1.2 \times 10^{40}$~erg~s$^{-1}$ (NGC~5005),
respectively.

\section{Comparison with ASCA/BeppoSAX results}

In this Section we compare the results of the
{\it Chandra} and XMM-Newton observations described
in Sect.~4 with prior ASCA and BeppoSAX measurements.
All the spectra described in this Section were
extracted from calibrated and linearized event lists
available in the public archive, and reanalyzed by us.
Whenever more than one observation was available for 
a given source, we have considered the latest (in
no case significant spectral variability was observed
across different ASCA/BeppoSAX observations,
with the only exception of NGC~1068 (Guainazzi et al. 2000b;
Colbert et al. 2002). This exception does not
substantially affect any of the results discussed in this
paper.

Variability in the soft (0.5--2~keV)
and hard (2--10~keV) X-ray flux is generally restricted
to a factor of $\pm$3
\cite{guainazzi04b}.
The {\it intensities} (but not the
{\it EWs}: see Sect.~6 below) of the K$_{\alpha}$ iron
lines are consistent with a
factor $\pm 2$ as well. The only exception is NGC~3393
(cf. Table~\ref{tab2}, and Table~\ref{tab3}),
where a delay in the response
of a variable primary continuum probably occurs.

In the following, some additional details are given
on the analysis of the ASCA/BeppoSAX observations,
whenever our analysis reaches
further or different conclusions with respect to what published
in the literature, or shown by the XMM-Newton observations.
\\[0.25cm]
{\bf NGC~5135}: the absorption-corrected 0.5-2~keV flux during
the January 1995 ASCA observation was $(1.2 \pm^{3.2}_{0.8}) \times
10^{-12}$~erg~cm$^{-2}$~s$^{-1}$. At face value this is one
order of magnitude larger than measured by {\it Chandra} 6.6 years
later. However, the difference is at the 1-$\sigma$ level only,
if the statistical uncertainties are taken into account.
The other spectral parameters are consistent with the {\it Chandra}
results, with large errors.
\\[0.25cm]
{\bf NGC~3393}: NGC~3393 is one of the few targets in our sample, which was
detected by the PDS instrument on board BeppoSAX above 15~keV
(count rate: $0.39 \pm 0.09$~s$^{-1}$). The BeppoSAX observation
is discussed by Maiolino et al. (1998). They interpret the
BeppoSAX spectrum as due to a Compton-thick source,
with a column density $> 10^{25}$~cm$^{-2}$.
The overall poor statistics of the BeppoSAX observation
prevented them from applying more complex models.
However, the PDS data points
in their Fig.~1 lay systematically above the extrapolation
of the best-fit model in the 2--10~keV band.

\begin{figure}
\psfig{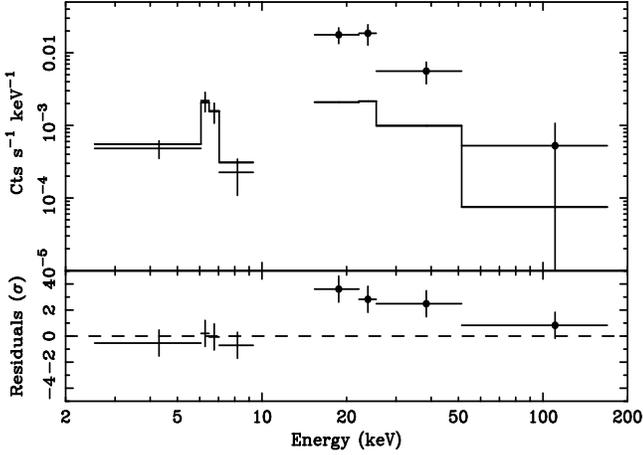}
\caption{Spectra ({\it upper panel}) and residuals in
units of standard deviations ({\it lower panel})
when the XMM-Newton best-fit spectrum
is applied (with $\Gamma \equiv 1.6$) to
the BeppoSAX spectra of NGC~3393 (MECS: {\it crosses};
PDS: {\it filled dots})}
\label{fig5}
\end{figure}

\begin{figure}
\psfig{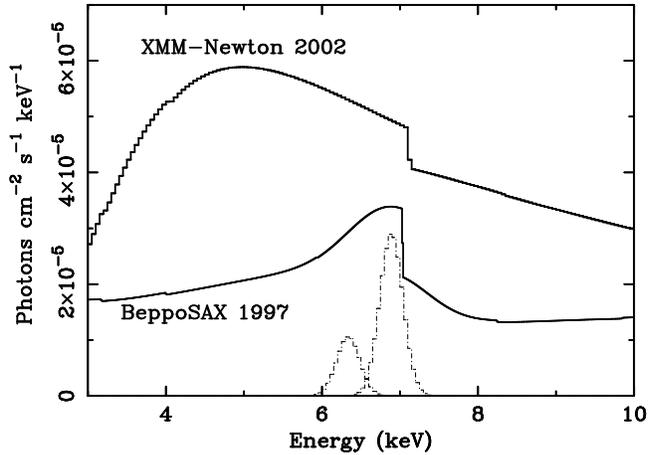}
\caption{3--10~keV total spectral energy distributions
of NGC~4939 based on the 1997 BeppoSAX and
2002 XMM-Newton observations ({\it solid lines}).
The {\it dot-dashed lines} indicate the upper limits
for the 6.4~keV and 6.96~keV K$_{\alpha}$ fluorescent
emission lines in the XMM-Newton observation.
For display purpose only, we have attributed to
all emission lines a formal width equal
to the intrinsic instrumental energy resolution:
150~eV and 500~eV for XMM-Newton (pn) and
BeppoSAX (MECS), respectively}
\label{fig13}
\end{figure}

\begin{figure}
\psfig{file=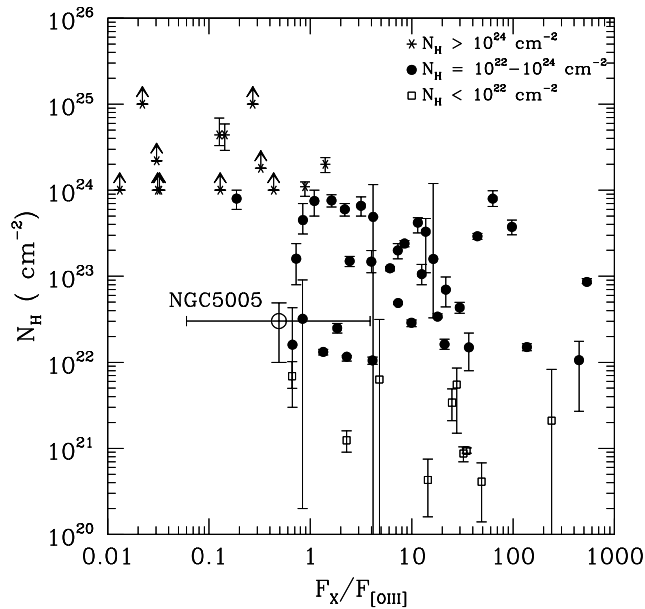,height=85mm,width=85mm}
\caption{Absorbing column density $N_H$
versus  2--10~keV to O[{\sc iii}] flux
ratio for the sample of Seyfert~2 galaxies
after Bassani et al. (1999). The {\it empty circle}
indicates the location corresponding to
NGC~5005 for the column density observed
by XMM-Newton, and the O[{\sc iii}] fluxes
reported in the literature (details in text)}
\label{fig8}
\end{figure}

We have first applied
the best model of the XMM-Newton observation
to the BeppoSAX spectra. As NGC~3393 is undetected
by the LECS instrument below 2~keV, we
kept the parameters of the thermal
components and of the $E_c \simeq 1.8$~keV
emission line frozen to the XMM-Newton
best-fit values, as the contribution of these
components is negligible in the MECS-PDS
energy bandpass.  The best-fit
intrinsic spectral
index face value turns out to be $\Gamma \simeq 0.7$. Although
the quality of the fit is acceptable ($\chi^2/\nu = 21.4/17$),
this flat index - still consistent with the XMM-Newton
results within the large statistical uncertainties - is suggestive
of additional spectral complexity. If
$\Gamma$ is fixed to the XMM-Newton best-fit value (1.6), the
PDS counts are largely underpredicted (Fig.~\ref{fig5}).
%---------------------------------------
%---------------------------------------
The difference is
even larger if more typical values
$\ge 1.9$ are used. We conclude that the flux in the
PDS band is dominated by the nuclear emission piercing
through a Compton-thick absorber, with $N_H < 10^{25}$~cm$^{-2}$.
Adding an absorbed power-law to the XMM-Newton best-fit
model
yields an improvement in the quality of the fit
at the
96.7\% confidence level ($\Delta \chi^2/\Delta \nu =
7.8/2$), with $N_H \sim 4 \times 10^{24}$~cm$^{-2}$,
and a slightly steeper intrinsic spectral index.
The best-fit parameters are shown in Table~\ref{tab3}.
These results confirm that
%---------------------------------------
\begin{table}
\caption{Best-fit parameters and results when
the XMM-Newton model (expanded with
a power-law covered by an absorber
of column density $N_H$ to account for
the emission in the PDS energy band)
is applied to the BeppoSAX
spectra of NGC~3393. Details in text.}
\begin{tabular}{lc} \hline \hline
$\Gamma$ & $2.8 \pm^{1.2}_{0.7}$ \\
$N_H$ ($10^{24}$~cm$^{-2}$) & $4.4 \pm^{2.5}_{1.1}$ \\
\multicolumn{2}{l}{Fe K$_{\alpha}$ line:} \\
$E_c$ (keV) & $6.58 \pm^{0.18}_{0.21}$ \\
$I_c$$^a$ & $1.4 \pm 0.8$ \\
$EW$ (keV) & $4 \pm 2$ \\ 
$\chi^2/\nu$ & 13.6/15 \\
\hline \hline
\end{tabular}

\noindent
$^a$in units of $10^{-5}$~photons~cm$^{-2}$

\label{tab3}
\end{table}
%---------------------------------------
the column density covering the NGC~3393 nucleus is indeed
Compton-thick, although not large enough to fully suppress
the nuclear emission.
\\[0.25cm]
{\bf NGC~4939}: NGC~4939 was classified as a Compton-reflection
dominated Compton-thick AGN by Maiolino et al. (1998),
on the basis of the very flat spectral index obtained
in the ``transmission-scenario", and the fact that
this model underpredicts the emission in the PDS band
(13-200~keV count rate: $0.20 \pm 0.06$~s$^{-1}$).
We have reanalyzed the same data, obtaining results which are
basically consistent with theirs. The EW of a single
Gaussian profile accounting for the observed iron
emission line ($\simeq 750$~eV) is indeed too large
with respect to the measured column density in
the transmission scenario ($N_H \simeq 1.3 \times 10^{23}$~cm$^{-2}$).
A fit where
hard X-rays are dominated by a ``bare" Compton-reflection
is excellent ($\chi^2/\nu = 45.4/55$; Table~\ref{tab4}).
%---------------------------------------
\begin{table}
\caption{Best-fit parameters and results
for the best-fit to the BeppoSAX spectra
of NGC~4939. Details in text.}
\begin{tabular}{lc} \hline \hline
$\Gamma_{hard}$ & $1.90 \pm^{0.16}_{0.19}$ \\
$\Gamma_{soft}$ & $3.5 \pm^{0.4}_{0.5}$ \\
$N_H$ ($10^{24}$~cm$^{-2}$)$^a$ & $>2$ \\
0.5--2~keV flux$^b$ & $0.43 \pm^{0.14}_{0.15}$ \\
2--10~keV flux$^b$ & $1.6 \pm 0.2$ \\
\multicolumn{2}{l}{Fe K$_{\alpha}$ lines:} \\
$I_{c,6.4}$$^c$ & $1.1 \pm 0.6$ \\
$EW_{c,6.4}$ (eV) & $490 \pm 270$ \\
$I_{c,6.96}$$^c$ & $1.2 \pm^{0.6}_{0.7}$ \\
$EW_{c,6.96}$ (eV) & $460 \pm^{270}_{230}$ \\
$\chi^2/\nu$ & 45.4/55 \\
\hline \hline
\end{tabular}

\noindent
$^a$assuming $\Gamma_{hard} \equiv 1.9$

\noindent
$^b$in units of $10^{-12}$~erg~cm$^{-2}$~s$^{-1}$

\noindent
$^c$in units of $10^{-5}$~photons~cm$^{-2}$

\label{tab4}
\end{table}
%---------------------------------------
The reflection-dominated state is confirmed by the large
EW of the neutral component of the iron K$_{\alpha}$ fluorescent
line ($EW \simeq 500$~eV). The H-like component exhibits
a comparable EW. The combination of hard X-ray continuum
and iron emission line EW points to a transition between
a ``reprocessing-" and a ``transmission-dominated" state
occurring
between the January 1997 BeppoSAX and the 
March 2001 XMM-Newton observation.
A comparison between the 3--10~keV spectral
energy distribution, based on the best-fit
BeppoSAX and XMM-Newton models, is shown in
Fig.~\ref{fig13}.
%---------------------------------------
%---------------------------------------
It is interesting to observe that the soft ($E \le 2$~keV) X-ray
flux {\it decreased} by a factor $\simeq$3.5
between the BeppoSAX and the XMM-Newton observation.
This supports our interpretation of the soft X-ray emission
in this object as due to scattering of the primary nuclear
continuum, which was mirroring a previous phase of
strong AGN activity during the BeppoSAX observation.
\\[0.25cm]
{\bf NGC~5005}:
NGC~5005 was declared Compton-thick by Risaliti et al.
(1999) on the basis of the low 
X-ray versus O[{\sc iii}] luminosity ratio, although no evidence for either
a flat hard X-ray spectrum or for a K$_{\alpha}$ iron
line was observed in the ASCA spectrum. The upper limit
of the EW of the latter feature (900~eV) was
still consistent
with heavy obscuration. None of the criteria adopted
to classify this source as a Compton-thick object
resists scrutiny after the XMM-Newton
observation. The upper limit on the K$_{\alpha}$ iron line
EW is strict ($< 240$~eV).
The application of the best-fit XMM-Newton model
(cf. Tab.~\ref{tab2}) to the ASCA
yields a good fit ($\chi^2/\nu=216.1/227$), showing that
the ASCA data have not enough statistics to
distinguish a ``transmission-~'' from a ``reprocessing-dominated''
scenario on the basis of the X-ray continuum shape.
Literature measurements of the O[{\sc iii}] flux - once
corrected for optical reddening using the prescription
in Bassani et al. (1999) - span
a rather large interval, between 0.3 and $20 \times
10^{-13}$~erg~cm$^{-2}$~s$^{-1}$ (Shuder \& Osterbrock 1981;
Dahari \& De Robertis 1988; Ho et al. 1997; Risaliti et al. 1999).
This interval is consistent with 2--10~keV
versus O[{\sc iii}] ratio values observed in
Compton-thin as well as Compton-thick Seyfert~2s (Fig.~\ref{fig8}).
%---------------------------------------
%---------------------------------------
We therefore conclude that NGC~5005 is
most likely a mis-classified Compton-thin
Seyfert~2.

\section{Discussion}

\subsection{How much do we know of Compton-thick Seyfert~2 galaxies?}

The safest way to identify a Compton-thick
Seyfert~2 galaxy, and describe - even at the
simplest phenomenological level - the basic
X-ray properties of its nuclear emission
is to detect the primary continuum
piercing through the Compton-thick absorber.
This requires measurement above 10~keV, which
have been possible so far only on the $\sim$10
objects detected by the PDS instrument on-board
BeppoSAX. For all the remaining known $\simeq$40
Compton-thick Seyfert~2
\cite{comastri04} the classification relies on
indirect evidence, such as the flatness of the
hard X-ray continuum, the EW of the
K$_{\alpha}$ iron fluorescent emission line(s),
or anomalous low values of the ratio between
the flux in the 2--10~keV energy band
and in other wavelengths.

Waiting for an X-ray detector of better $>10$~keV
sensitivity
than the PDS, the robustness of the criteria
used to identify Compton-thick objects can be
tested with the improved sensitivity that
the XMM-Newton optics offer. In our sample, the Compton-thick
nature is confirmed for all objects, except
NGC~5005 ($N_H \simeq 1.5 \times 10^{23}$~cm$^{-2}$),
and - marginally - NGC~5643 (Guainazzi et al. 2004a;
$N_H = 6$--$10 \times 10^{23}$~cm$^{-2}$),
apart from NGC~4939, obviously. This
is a potentially important result,
as it underlines the perspective to
extend the search for Compton-thick
objects at higher redshift \cite{fabian02}.
Although classification of an individual object
may be subject to uncertainties even when large
EW iron lines are detected, the method is
robust.

\subsection{The moderately unstable temper of
heavily obscured AGN}

The scope of this paper is comparing the X-ray
spectral properties of a complete, unbiased sample
of Compton-thick Seyfert~2 galaxies
observed with {\it Chandra} and
XMM-Newton with prior measurements.
The main scientific goal is to estimate the
rate of transitions between ``transmission-~''
and ``reprocessing-dominated'' spectral states.

These transitions were serendipitously discovered
on a few nearby active nuclei, once a larger database
of X-ray observations allowed us some knowledge of
the X-ray history of a wider sample of AGN. These
transitions affect Seyfert~2 galaxies (NGC~2992; Gilli
et al. 2000; NGC~1365; Iyomoto et al. 1997, Risaliti
et al. 2000; UGC~4203, Guainazzi et al. 2001; NGC~6300
Guainazzi 2002), as well as other
AGN (NGC~4051; Guainazzi et al. 1998,
Uttley et al. 1999; PG~2112+059; Gallagher et al. 2004).
It has been claimed that variability of the
absorbing column density by a factor $50 \pm 30\%$
on timescales $\le 1$~year is common in obscured AGN
\cite{risaliti02}. In one of the best-monitored cases
(NGC~3227; Lamer et al. 2003), the symmetric profile
of the absorption light curve
clearly suggests an interpretation
in terms of line-of-sight crossing by an individual cloud.
The transitions we are discussing in this paper, however,
represent a different phenomenology, whereby the apparent
variation of the absorbing column density is of
at least one order-of-magnitude, and
the state corresponding to the lower X-ray flux is
fully reprocessing-dominated.

\begin{figure}
\psfig{file=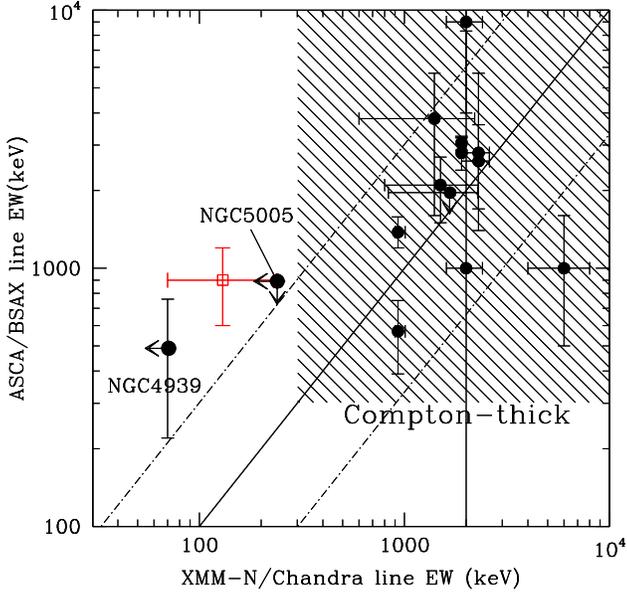,height=85mm,width=85mm}
\caption{EW of the K$_{\alpha}$ neutral
iron fluorescent emission line measured by
ASCA/BeppoSAX as a function of the
same quantity measured by {\it Chandra}/XMM-Newton.
The locus corresponding to ``static'' Compton-thick
objects is indicated by the {\it shaded area}.
{\it Filled circles}: objects of our sample;
{\it open square}: UGC~4203, the prototypical
``Phoenix Galaxy'' (Guainazzi et al. 2002).
The {\it solid} ({\it dashed}) line
represents the identity ($\pm$factor 3)
locus}
\label{fig9}
\end{figure}

The main conclusion of this study is summarized in
Fig.~\ref{fig9}, where we use the K$_{\alpha}$ iron
line EW - measured by BeppoSAX/ASCA and {\it Chandra}/XMM-Newton -
as a hallmark for {\it bona fide} Compton-thick AGN.
Only ``narrow'' or ``unresolved'' components
corresponding to neutral or mildly ionized
iron transitions have been used in the calculation of
the EWs shown in Fig.~\ref{fig9}. We do not consider
%---------------------------------------
%---------------------------------------
the contribution of Compton-shoulder, which is likely
to be $\approxlt 20\%$ in most cases \cite{matt02}.
The locus corresponding to ``Compton-thick'' objects
is conservatively bordered by the line
$EW \equiv 300$~eV, corresponding to
the brightest emission line produced by transmission through
a Compton-thin screen, covering a 2$\pi$ solid angle
\cite{leahy93}. Out of the 10 Compton-thick Seyferts
discussed in this paper, we find evidence for
only one transition: NGC~4939. This
implies that a typical timescale for these transitions
should be $\simeq$50~years, on the basis of
the average separation between the ASCA/BeppoSAX
and the closest {\it Chandra}/XMM-Newton
observations. Of course, given the
size of our sample, this has
to be regarded as no more than an order-of-magnitude
estimate. However, it is consistent with previous
determinations, based on inhomogeneous and incomplete
samples.

The origin of the spectral changes occurring when an
AGN transforms its appearance from a ``transmission-''
to a ``reprocessing-dominated'' state is still
not fully elucidated. In principle, high-quality,
high-resolution measurements, following the onset
of the variability or the AGN recovery after a
prolonged ``off-state'' should be decisive.

In NGC~6300 and NGC~2992
we have the strongest evidence that
transitions from
transmission- to reprocessing-dominated
states are due to a change of the optical path
through which the nucleus is being observed,
due to a temporary interruption of the nuclear
activity. In NGC~6300 this is suggested
by the very large Compton-reflection continuum
observed by
BeppoSAX \cite{guainazzi02}. In NGC~2992,
the history of the X-ray emission
is comparatively well sampled. The X-ray
flux decreased by about a factor of 20 from
the first HEAO-1 detection in the early 80s,
up to the 1994 ASCA observation, just to experience
a factor $\simeq$15 recovery by 1999
\cite{gilli00}. 
To these two cases, we might add the large
dynamical range of the
NGC~5643 AGN X-ray output \cite{guainazzi04b},
a promising transition candidate.
Unfortunately our
knowledge of the X-ray history
of NGC~4939, and UGC~4203 is too poor,
for us to be able to draw any final conclusions
on this issue. Recently Ohno et al.
(2004) proposed a change by a factor $>5$
of the absorber column density as the most
likely explanation in the latter. On the other hand,
multiple X-ray observations of
optically defined samples of unobscured AGN have
not detected a significant rate of large
historical X-ray flux variations 
(compare, for instance, Laor et al. 1997,
George et al. 1998, and Porquet et al. 2004).
When one is dealing with X-ray unobscured
AGN a bias toward brightest,
less obscured object, may prevent us from
discovering the fraction of X-ray unobscured
counterparts to our ``transient" Compton-thick Seyferts.
However, this may as well suggest
an alternative
interpretation in terms of changes in
the properties of line-of-sight
gas. This
scenario will be investigated in the next
Section.

\subsection{The recovery of fossil AGN as probe of
the circumnuclear medium}

\begin{figure}
\psfig{file=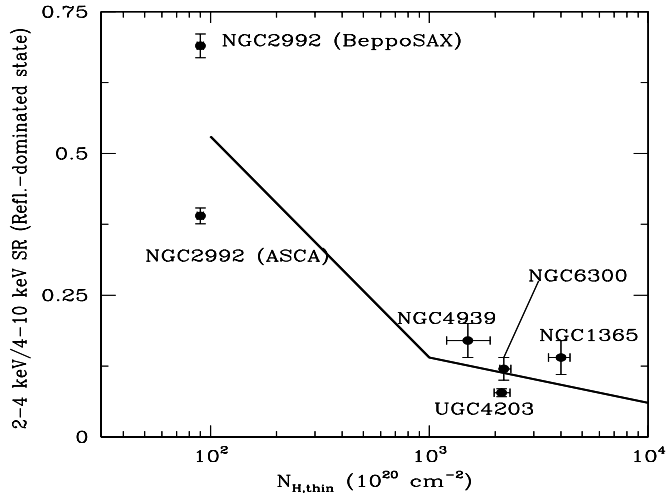,height=70mm,width=90mm}
\caption{2--4~keV to 4--10~keV flux softness ratio (SR)
in the reprocessing-dominated state against
the $N_H$ measured in transmission-dominated
states for the 5 objects, where a transition
between the two states has been reported in the literature
(cf. Sectc.~6.2), plus NGC~4939 (this paper).
The {\it solid line} represents the
expected correlation on the basis of
Monte-Carlo simulations of Compton-reflection
spectra (Matt et al. 2003b)}
\label{fig10}
\end{figure}

``Transmission-" to
``reprocessing-dominated" transitions can be
used to probe some properties of the
gas in the nuclear environment
in highly obscured AGN. The method - based on
Monte-Carlo simulations - is described by Matt et al. (2003b).
The application of this method to the five known
transitions is summarized in Fig.~\ref{fig10},
%---------------------------------------
%---------------------------------------
where we show the 2--4~keV versus 4--10~keV flux
softness ratio for
the Compton-reflection component
in the reprocessing-dominated state against the measured
column density in the transmission-dominated state.
The {\it solid line} represents the expected correlation
according to simulations. In the ASCA observation of NGC~2992,
and possibly in UGC~4203, the spectrum observed
during a reprocessing-dominated state is too hard
to be due to Compton-reflection by matter
with the same column density as measured
by the soft photoelectric cut-off during
transmission-dominated states. A similar
argument can be applied to NGC~6300, on
the basis of its X-ray spectrum above
10~keV (Guainazzi 2002; Matt et al. 2003b). The
gas responsible for line-of-sight absorption should be
therefore different in density from the gas
responsible for reflection, the latter
being most likely located
at the far inner side of the molecular ``torus''.
This may be explained by a largely inhomogeneous
compact ({\it i.e.}, 1~pc)
nuclear absorber \cite{ohno04},
or by Compton-thin absorption occurring
on much larger scales than the nuclear
``torus'', {\it e.g.} in dusty regions associated
with starburst formation \cite{weaver01},
or with the host galaxy (Maiolino \& Rieke 1995;
Malkan et al. 1998). An extension of the
Seyfert Unified scenario, which encompasses the
latter possibility is discussed by Matt (2000).
It is also noteworthy that in 3 out of 5 cases
the SR in Fig.~\ref{fig10} is softer then
expected by a pure Compton-reflection spectrum
even in the reprocessing-dominated states.
This suggest a still not-negligible contamination
by a softer component, {\it e.g.} a not
fully ``switched-off'' AGN continuum. This
possibility is confirmed by a detailed spectral analysis
of the BeppoSAX observation of NGC~2992, the X-ray brightest
among the objects displayed in Fig.~\ref{fig10}, where
the recovery of the AGN is witnessed by a comparatively
small value of the normalization ratio between the
reflected and the transmitted component [$R \simeq
4 (\Omega/2 \pi$)] with respect to a bare Compton-reflection
dominated state.

In summary,
the spectral properties of
``transmission-" to ``reprocessing-dominated"
transitions
indicate that obscuring matter in AGN is
far from being homogeneous in space or time.
Yet, one cannot discard
the idea of a compact
but inhomogeneous pc-scale ``torus" 
\cite{antonucci93}, or disk
outflow \cite{elvis00}. Such
inhomogeneities might be the ultimate responsible
for these transitions.
However, the
fact that photoionized- or starburst-dominated
soft X-ray emission in several Seyfert~2 galaxies
is absorbed as well (Iwasawa \& Comastri 1998;
Matt et al. 2001; 2003a), alongside
our knowledge of the X-ray history in
NGC~6300 and, above all, NGC~2992, suggests that an
important contribution to X-ray obscuration comes
from matter associated to the host galaxy,
beyond the innermost pc around the central
engine.

\subsection{The origin of the soft X-ray emission}

\begin{figure}
\psfig{file=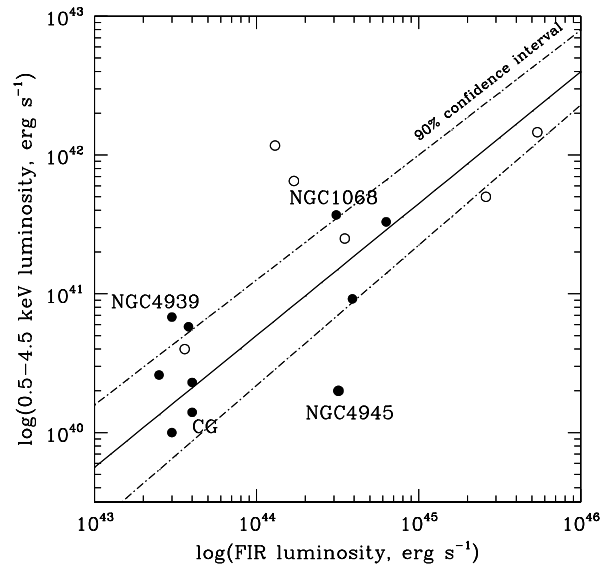,height=80mm,width=80mm}
\caption{FIR luminosity (as defined in
David et al. 1992) against the 0.5--4.5~keV
luminosity for the Compton-thick objects
of our sample ({\it filled dots}),
and the remaining Compton-thick
Seyfert~2 galaxies of Bassani et al. (1999).
The {\it solid line} indicates
the empirical correlation found
by David et al. (1992) for
starburst galaxies; the {\it
dashed lines} indicates the 90\% confidence
level on this correlation. `CG'': the
Circinus Galaxy}
\label{fig11}
\end{figure}

Our program was not specifically tuned to investigate
the origin of the soft X-ray emission in our sample.
The answers that we get from the data on this issue
are therefore
inevitably ambiguous in almost all cases. The only
exception is NGC~5005
(the only non Compton-thick source in
the sample). In this case, the soft X-ray emission
is clearly extended on scales comparable with
the optical size of the galaxy, and roughly aligned
with its main axis, or with an inner arm UV structure.
For all the other cases, the two proposed scenarios
are equally viable on the basis of the EPIC data alone.
Whenever high-resolution spectroscopic data are available,
soft X-rays appear to be dominated by emission lines
following photoionization and photoexcitation by the
primary AGN emission (Kinkhabwala et al. 2002;
Sambruna et al. 2001; Bianchi et al. 2001), with
little or no contribution by nuclear starbursts.
On the other hand, the 0.5--4.5~keV X-ray
luminosities are generally
consistent with the correlation with
the Far InfraRed (FIR) luminosity empirically determined
by David et al. (1992) on a large sample of starburst-dominated
galaxies (Turner et al. 1997; Maiolino et al. 1998;
Fig.~\ref{fig11}), the only discrepant objects
being NGC~4939, and NGC~4945.
%---------------------------------------
%---------------------------------------
This, however, holds as well to objects
(such as the Circinus Galaxy), for which it
is unlikely that soft X-rays are dominated by starburst.
On the other hand, high-resolution
imaging of NGC~4945 with {\it Chandra}
shows that soft X-rays are likely to be dominated
by thermal emission from starbursts, alongside
a starburst mass-loaded superwind \cite{schurch02}.
While we refer to Guainazzi et al. (2004a)
for a more detailed discussion on this point, we
conclude for the time being that
a throughout determination of the physical properties
of the plasma dominating the soft X-rays in obscured
AGN requires deep exposures with high-resolution
detector, which are currently possible only on
a limited number  of objects.

\section{Conclusions}

The main result of this paper is the
estimate of the occurrence rate
of ``transmission-" to ``reprocessing-dominated"
state transitions, on the most unbiased and complete
existing sample of Compton-thick Seyfert~2 galaxies
(cf. Fig.~\ref{fig9}). We have discovered 1
(new) transition
out of a sample of 10 {\it bona-fide}
Compton-thick objects, with an average
time span between pre- and post XMM-Newton and {\it
Chandra} launch observations of about 5~years.
The statistics is still too small to determine
anything more accurate than
an order-of-magnitude estimate for the
occurrence rate. Bearing this caveat in mind,
it seems that once or twice every century we might
be forced to change our X-ray absorption classification
for each highly obscured AGN.

With respect to
the mechanism responsible for these transitions,
our results are consistent with the discussion in
Matt et al. (2003b). Although an explanation
in terms
of varying line-of-sight column density cannot be ruled
out, the fact that in the best studied cases these
transitions are associated with large ($>$10)
fluctuations of the AGN X-ray
output suggests that they are
due to a change of the optical path through which
we observe the nucleus. For the transition
specifically discussed in this paper - NGC~4939 -
the factor of 2 variation of the
observed 2--10~keV band flux hides
a larger variation of the AGN intrinsic
power, as its true luminosity in
the reprocessing-dominated state is unknown.
If this is the correct
interpretation, the transition occurrence rate
translates immediately into a duty-cycle of the
AGN phenomenon in the local universe.

Independently of the ultimate mechanism responsible
for these transitions, comparison of their
spectral properties with Monte-Carlo simulations
demonstrates that obscuring gas in absorbed
AGN cannot be distributed in a space- or time- homogeneous
structure. Again, a compact but inhomogeneous ``torus" cannot
be ruled out. However, there is mounting evidence that
gas in regions of intense star formation and
dust in the host galaxy play a major role, and
might be ultimately responsible for the bulk
of Compton-thin X-ray absorption in AGN.

\section*{Acknowledgments}

This paper is based on observations obtained with XMM-Newton, an ESA
science mission with instruments and contributions directly funded by
ESA Member States and the USA (NASA). This research has made use of
data obtained through the High Energy Astrophysics Science Archive
Research Center Online Service, provided by the NASA/Goddard Space
Flight Center and of the NASA/IPAC Extragalactic Database (NED) which
is operated by the Jet Propulsion Laboratory, California Institute of
Technology, under contract with the National Aeronautics and Space
Administration. GM acknowledges financial support from MIUR grant
COFIN-03-02-23. Careful reading, and insightful suggestions by
an anonymous referee are gratefully acknowledged.

{}

\end{document}